\documentclass[aps,prl,10pt,twocolumn,superscriptaddress,floatfix,nofootinbib,amsmath,amssymb,altaffilletter,preprintnumbers]{revtex4-1}
\usepackage[colorlinks=true,pdfstartview=FitV,breaklinks=true]{hyperref}

\usepackage{bm}
\usepackage{graphicx}
\usepackage{multirow}
\usepackage{soul}
\usepackage{amsmath}
\usepackage{fontawesome}
\usepackage{mathrsfs}
\usepackage{amssymb}
\usepackage{yfonts}
\usepackage{makecell}
\usepackage{color}
\usepackage{xspace}
\usepackage{url}
\usepackage{verbatim}
\usepackage{mathtools}
\usepackage{upgreek}
\usepackage{units}
\usepackage{siunitx}
\usepackage{amstext}
\usepackage{times}
\usepackage{booktabs}
\usepackage{tabulary}
\usepackage{tabularx}
\usepackage{etoolbox}
\usepackage[version=4]{mhchem}
\usepackage[utf8]{inputenc}
\usepackage[dvipsnames,table]{xcolor}
\newcolumntype{Y}{>{\centering\arraybackslash}X}
\hypersetup{urlcolor=BlueViolet,
	    citecolor=Plum,
	    linkcolor=PineGreen}
\usepackage[official]{eurosym}
\definecolor{lightgray}{rgb}{0.9,0.9,0.9}	    
\definecolor{green}{rgb}{0,0.5,0}
\definecolor{red}{rgb}{1,0,0}
\definecolor{blue}{rgb}{0,0,0.5}

\long\def\symbolfootnote[#1]#2{\begingroup%
\def\thefootnote{\fnsymbol{footnote}}\footnotetext[#1]{#2}\footnotemark[#1]\endgroup}

\newcommand{\vmin}{v_{\rm min}}

\newcommand{\dbd}[2]{\ifmmode \frac{\textrm{d}#1}{\textrm{d}#2}\else $\textrm{d}#1/\textrm{d}#2$\fi}
\newcommand{\pbp}[2]{\ifmmode \frac{\partial#1}{\partial#2}\else $\partial#1/\partial#2$\fi}

\newcommand{\drm}{\mathrm{d}}

\DeclareMathAlphabet{\mathpzc}{OT1}{pzc}{m}{it}
 \newcommand{\eV}{\text{e\kern-0.15ex V}\xspace}

 \newcommand{\TeV}{\text{T\kern-0.1ex \eV}\xspace}
 
 \newcommand{\cevns}{CE$\nu$NS\xspace}

\newcommand{\Boron}{$^8$B\xspace}

\DeclareMathAlphabet{\mathpzc}{OT1}{pzc}{m}{it}

\newcommand{\be}{\begin{equation}}
\newcommand{\ee}{\end{equation}}
\newcommand{\bea}{\begin{eqnarray}}
\newcommand{\eea}{\end{eqnarray}}

\begin{document}

\title{Fog on the horizon: a new definition of the neutrino floor for direct dark matter searches}

\author{Ciaran A. J. O'Hare}\email{ciaran.ohare@sydney.edu.au}
\affiliation{School of Physics, The University of Sydney, and ARC Centre of Excellence for Dark Matter Particle Physics, NSW 2006, Camperdown, Sydney, Australia}

\smallskip
\begin{abstract}
The neutrino floor is a theoretical lower limit on WIMP-like dark matter models that are discoverable in direct detection experiments. It is commonly interpreted as the point at which dark matter signals become hidden underneath a remarkably similar-looking background from neutrinos. However, it has been known for some time that the neutrino floor is not a hard limit, but can be pushed past with sufficient statistics. As a consequence, some have recently advocated for calling it the ``neutrino fog'' instead. The downside of current methods of deriving the neutrino floor are that they rely on arbitrary choices of experimental exposure and energy threshold. Here we propose to define the neutrino floor as the boundary of the neutrino fog, and develop a calculation free from these assumptions. The technique is based on the derivative of a hypothetical experimental discovery limit as a function of exposure, and leads to a neutrino floor that is only influenced by the systematic uncertainties on the neutrino flux normalisations. Our floor is broadly similar to those found in the literature, but differs by almost an order of magnitude in the sub-GeV range, and above 20~GeV.\, \href{https://github.com/cajohare/NeutrinoFog}{\large\faGithub}
\end{abstract}

\maketitle

\begin{figure}[htb]
\begin{center}
\includegraphics[trim = 0mm 0mm 0mm 0mm, clip, width=0.49\textwidth]{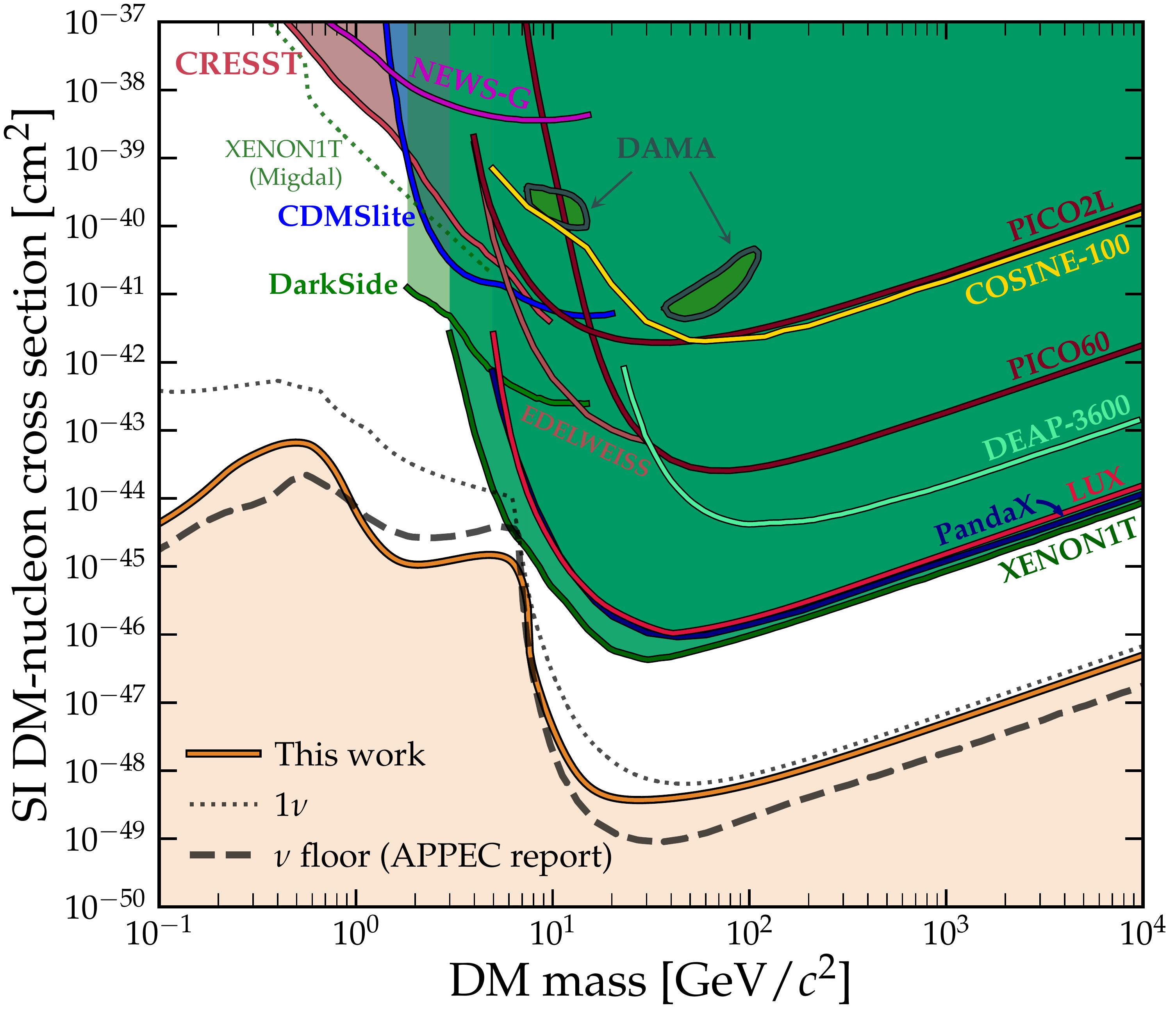}
\caption{Present exclusion limits on the spin-independent DM-nucleon cross section (assuming equal proton/neutron couplings)~\cite{CDMSlite,Adhikari:2018ljm,CRESST:2019jnq,Bernabei:2018yyw,Savage:2008er,DarkSide:2018bpj,DEAP:2019yzn,EDELWEISS:2016nzl,LUX:2016ggv,NEWS-G:2017pxg,PandaX-II:2017hlx,Amole:2016pye, Amole:2017dex,XENON:2019zpr,XENON:2020gfr}. Beneath these limits we show three definitions of the neutrino floor for a xenon target. The previous discovery-limit-based neutrino floor calculation shown by the dashed line is taken from the recent APPEC report~\cite{Billard:2021uyg} (based on the technique of Ref.~\cite{Billard:2013qya}). The envelope of 90\% C.L.~exclusion limits seeing one expected neutrino event is shown as a dotted line. The result of our work is the solid orange line. We define this notion of the neutrino floor to be the boundary of the neutrino fog, i.e.~the cross section at which any experiment sensitive to a given value of $m_\chi$ leaves the standard Poissonian regime $\sigma \propto N^{-1/2}$, and begins to be saturated by the background: $\sigma \propto N^{-1/n}$, with $n>2$. The floor is thus a contour for $n=2$.} 
\label{fig:NewFloor}
\end{center}
\end{figure} 

\begin{figure*}
\begin{center}
\includegraphics[trim = 0mm 0mm 0mm 0mm, clip, width=0.98\textwidth]{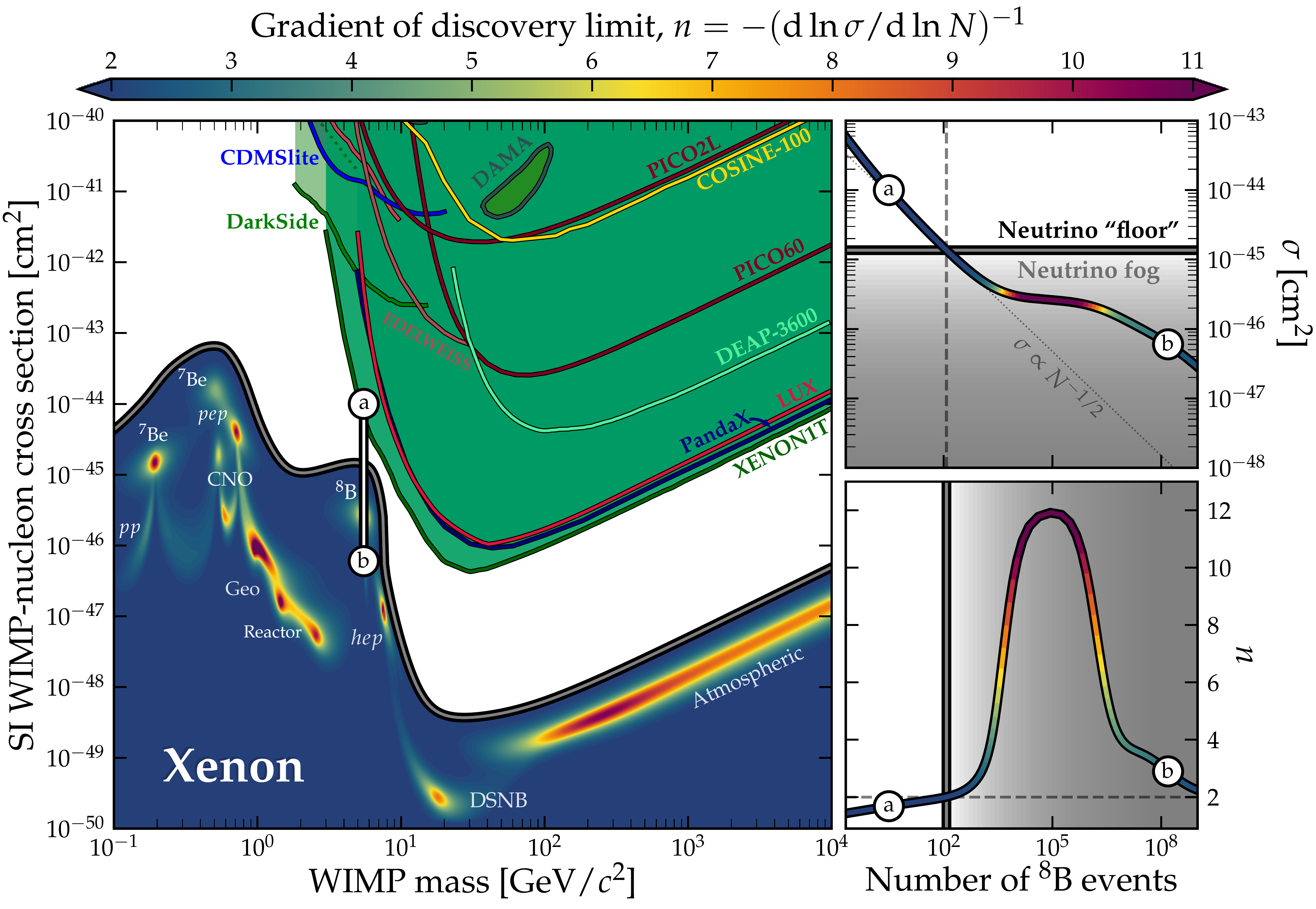}
\caption{A graphical description of the technique we adopt to map the neutrino fog and plot its boundary. In the main panel we show the spin-independent DM parameter space, colouring the section below the neutrino floor by the value of $n$, defined as the index with which a discovery limit scales with the number of background events, i.e. $\sigma \propto N^{-1/n}$. The neutrino fog is defined to be the regime for which $n>2$, with the neutrino floor being the cross section for a given mass where this transition occurs. The top right panel shows the evolution of $\sigma$ with $N$ at $m_\chi = 5.5$ GeV between the two cross sections labelled ``a'' and ``b'' on the main panel. The lower right panel shows the value $n$, found from derivative of the curve in the top right panel.} 
\label{fig:NuFloorExplained}
\end{center}
\end{figure*}

\textit{\textbf{Introduction}}.---Modern experiments searching for dark matter (DM) in the form of weakly interacting massive particles (WIMPs) have become rather large~\cite{Battaglieri:2017aum,Schumann:2019eaa}. It has been anticipated for some time~\cite{Monroe:2007xp,Vergados:2008jp,Strigari:2009bq,Gutlein:2010tq} that these underground detectors might one day become large enough to detect not just DM, but astrophysical neutrinos as well. In fact, it appears as though the first detection of solar neutrinos in a xenon-based detector is just around the corner~\cite{XENON:2020gfr}. Fittingly, the community has begun to collate a rich catalogue of novel physics to be done with our expanding multi-purpose network of large underground detectors~\cite{Harnik:2012ni,Pospelov:2011ha,Billard:2014yka,Franco:2015pha,Schumann:2015cpa,Strigari:2016ztv,Dent:2016wcr,Chen:2016eab,Cerdeno:2016sfi,Dutta:2019oaj,Lang:2016zhv,Bertuzzo:2017tuf,Dutta:2017nht,Leyton:2017tza,AristizabalSierra:2017joc,Boehm:2018sux,Bell:2019egg,Newstead:2018muu,DARWIN:2020bnc,LZ:2021xov}.

Unfortunately for WIMP enthusiasts, the impending arrival of neutrinos in DM detectors is somewhat bittersweet---being, as they are, essentially the harbingers of the end of conventional searches. These experiments usually look for signals of DM using nuclear recoils---a channel through which neutrinos also generate events via coherent elastic neutrino-nucleus scattering (\cevns)~\cite{Freedman:1973yd,Freedman:1977,Drukier:1983gj}. It turns out that the recoil signatures of DM and neutrinos look remarkably alike, with different sources of neutrino each masquerading as DM of varying masses and cross sections~\cite{Billard:2013cxa}.

Even an irreducible background like neutrinos may not be so problematic were it not for the---sometimes sizeable---systematic uncertainties on their fluxes. The cross section below which the potential discovery of a DM signal is prohibited due to this uncertainty is what is usually, but not always, labelled the ``neutrino floor''~\cite{Billard:2013qya}: a limit that has since been the subject of many detailed studies~\cite{OHare:2016pjy,Dent:2016iht, Dent:2016wor, Gelmini:2018ogy,AristizabalSierra:2017joc,Gonzalez-Garcia:2018dep,Papoulias:2018uzy,Essig:2018tss,Wyenberg:2018eyv,Boehm:2018sux,Nikolic:2020fom,Munoz:2021sad,Calabrese:2021zfq,Sierra:2021axk}. Since 2013 some form of neutrino floor has been shown underneath all experimental results, often billed as an ultimate sensitivity limit~\cite{Gibney:2020}. Methods of circumventing the neutrino floor have been proposed~\cite{Ruppin:2014bra,Davis:2014ama,Sassi:2021umf}. However, only directional detection seems to be a realistic strategy for doing so with comparatively low statistics~\cite{O'Hare:2015mda,Grothaus:2014hja,Mayet:2016zxu,OHare:2017rag,Franarin:2016ppr,OHare:2020lva,Vahsen:2020pzb,Vahsen:2021gnb}.

One potentially misleading aspect of the name ``neutrino floor'' is the fact that while it does pose an existential threat to DM searches, the floor itself is not solid. Firstly, the severity of the neutrino background---and hence the height of the floor in terms of cross section---is dependent crucially on neutrino flux uncertainties, which are anticipated to improve over time. Secondly, the DM and neutrino signals are never perfect matches. Even for DM masses and neutrino fluxes with very closely aligned nuclear recoil spectra---like xenon scattering with $^8$B neutrinos and a 6 GeV WIMP---they are not precisely the same. This means that with a large enough number of events, the spectra should be distinguishable~\cite{Ruppin:2014bra}. This fact implies that neutrinos present not a floor, but perhaps a `fog': a region of the parameter space where a clear distinction between signal and background is challenging, but not impossible.

The fogginess of the neutrino floor has become somewhat better appreciated recently~\cite{Gelmini:2018ogy,Sassi:2021umf,Gaspert:2021gyj}. However it is something that is rarely visualised: usually just a single neutrino floor limit is plotted. The most common version shown relies on an interpolation of several discovery limits for a set of somewhat arbitrary thresholds and exposures. It is true that given the softness of the neutrino fog, insisting upon a hard boundary will always be slightly arbitrary, however it should be possible to devise a simpler and self-consistent definition.

Since direct detection experiments will venture into the neutrino fog imminently, it is timely to update and refine our definitions. In this Letter we propose a new definition of the neutrino floor that situates it at the edge of the neutrino fog. We aim for this definition to 1) not depend upon arbitrary choices for experimental thresholds, or absolute numbers of observed neutrino events, 2) have a single consistent statistical meaning, and 3) be flexible to future improvements to neutrino flux measurements. The result of this effort can be found in Fig.~\ref{fig:NewFloor}, contrasted against previously used definitions. The technique for calculating this limit is explained graphically in Fig.~\ref{fig:NuFloorExplained}. The rest of this paper is devoted to summarising the basic ingredients of the calculation and evaluating several illustrative examples to compare against previous versions.

\textit{\textbf{Neutrinos versus DM}}.---To begin, we need to define a DM model to frame our discussion around. We adopt the following DM-nucleus scattering rate,
\begin{equation}\label{eq:wimprate}
   \frac{\drm R_\chi}{\drm E_r} = \rho_0 \frac{\mathcal{C}\sigma}{2 m_\chi \mu^2} F^2(E_r) \, g(\vmin) \, ,
  \end{equation}
where $\rho_0$ is the local DM density, $m_\chi$ is the DM mass, $\sigma$ is some DM-nucleon cross section, $\mu$ is the DM-nucleon reduced mass, $\mathcal{C}$ is a nucleus-dependent constant that coherently enhances the rate, and $F(E_r)$ is the form factor that suppresses it at high energies. Finally, $g(v_{\rm min})$ is the mean inverse DM speed above the minimal speed required to produce a recoil with energy $E_r$. The latter is found by integrating the lab-frame DM velocity distribution. We assume the time-averaged Standard Halo Model, with parameters summarised in Ref.~\cite{Evans:2018bqy}. Alternative halo models will lead to different neutrino floors, and including the unknown velocity distribution as an additional systematic uncertainty will act to raise the neutrino floor in general~\cite{OHare:2016pjy}. For this initial explanation we will frame the discussion around the spin-independent (SI) isospin-conserving DM-nucleon cross section $\sigma \equiv \sigma^{\rm SI}_p$, which is the canonical cross section that experimental collaborations most frequently set exclusion limits on. The scattering rate for this model is enhanced by $\mathcal{C} = A^2$, for a target with $A$ nucleons.

Neutrinos can scatter elastically off nuclei and produce recoils with very similar spectra to the ones found by evaluating Eq.(\ref{eq:wimprate}). Currently, the only measurement of \cevns is by COHERENT~\cite{Akimov:2017ade, Akimov:2020pdx}, but it is well-understood in the Standard Model~\cite{Freedman:1973yd,Freedman:1977}. Theoretical uncertainties, for example from the running of the Weinberg angle~\cite{Erler:2004in}, or the nuclear form factor~\cite{Papoulias:2018uzy}, are subdominant to those on the neutrino fluxes (see for instance Ref.~\cite{Sierra:2021axk}). We assume a Weinberg angle of $\sin^2{\theta_W} = 0.2387$, and for both neutrino and SI DM scattering we use the standard Helm form factor~\cite{Lewin:1995rx}. Summaries of the calculation of the \cevns cross section $\mathrm{d}\sigma_{\nu N}(E_\nu)/\mathrm{d}E_r$ as a function of neutrino and recoil energy, as well the resulting spectra of recoil energies for the various targets we will consider here can be found in e.g.~Refs.~\cite{Scholberg:2005qs,Billard:2013qya,Ruppin:2014bra,OHare:2020lva}.

The recoil energy spectra are found by integrating the differential \cevns cross section multiplied by the neutrino flux,
\begin{equation}
   \dbd{R_\nu}{E_r}= \frac{1}{m_N} \int_{E_{\nu}^{\min }} \dbd{\Phi}{E_\nu} \dbd{\sigma_{\nu N}(E_\nu)}{E_r} \mathrm{d}E_{\nu} \, .
\end{equation}
We cut off the integral at the minimum neutrino energy that can cause a recoil with $E_r$: $E_{\nu}^{\min }=\sqrt{m_{N} E_{r} / 2}$. We adopt the same neutrino flux model as in Ref.~\cite{OHare:2020lva} (Table I), so we will only briefly summarise some pertinent details. Further information about neutrino fluxes can be found in Ref.~\cite{Vitagliano:2019yzm}.

\emph{Solar neutrinos} generated in nuclear fusion reactions in the Sun form the largest flux at Earth for $E_\nu \lesssim 11$~MeV. These will be the primary source of \cevns events for most DM detectors and will limit discovery around $m_\chi\sim10$~GeV. The Sun's nuclear energy generation is well-understood, and in the case of the most important flux of neutrino from $^8$B decay, the corresponding flux normalisation, is also measured precisely~\cite{Super-Kamiokande:2001ljr,Borexino:2008fkj,Borexino:2017uhp,Super-Kamiokande:2010tar,KamLAND:2011fld,SNO:2011hxd,SNO:2018fch}. For the less well-measured components, several theoretical calculations of the solar neutrino fluxes are on the market (see e.g.~Ref.~\cite{Gann:2021ndb} for a recent review). Here we adopt the Barcelona 2016 calculation of the GS98 high-metallicity Standard Solar Model~\cite{Vinyoles:2016djt}. We adopt the quoted uncertainty on each flux normalisation, with the exception of $^8$B which we give a 2\% uncertainty in line with global fits of neutrino data~\cite{Bergstrom:2016cbh}. After $^8$B neutrinos, the most important solar fluxes are the two neutrino lines from electron capture by $^7$Be, which mimic the signal for sub-GeV masses---these come with 6\% uncertainties.

\emph{Geoneutrinos} are a constant flux of antineutrinos produced in radioactive decays of mainly uranium, thorium and potassium in the Earth. As might be expected, the ratios and normalisations of these fluxes are location-dependent. For concreteness, we use spectra from Ref.~\cite{Ludhova:2013hna} and normalise our geoneutrino fluxes to Gran Sasso, with corresponding uncertainties ranging from 17--25\%~\cite{Huang:2013}. Geoneutrinos impede the discovery at $m_\chi\sim$~GeV, but only for cross sections $\sigma^{\rm SI}_p \lesssim 10^{-47}$~cm$^2$.

\emph{Nuclear reactors} generate another source of antineutrinos and influence the floor at slightly higher masses. We assume the fission fractions and average energy releases from Ref.~\cite{Ma:2012bm} combined with the spectra from Ref.~\cite{Mueller:2011nm} before summing over all nearby nuclear reactors to Gran Sasso~\cite{Baldoncini:2014vda}.

\emph{The diffuse supernova neutrino background} (DSNB) is the cumulative flux of neutrinos from the cosmological history of core-collapse supernovae, and is relevant for the neutrino floor in a small mass window around 20 GeV. We adopt the fluxes parameterised in terms of three effective neutrino temperatures for the different flavour contributions, and place a 50\% uncertainty on the all-flavour flux~\cite{Beacom:2010kk}.

\emph{Atmospheric neutrinos} originate from the scattering of high-energy cosmic rays. The low-energy tail of the flux is small $\Phi\sim 10$ cm$^{-2}$~s$^{-1}$, but is the dominant background at high recoil energies. We use the theoretical flux model for 13 MeV--1 GeV atmospheric neutrinos from FLUKA simulations~\cite{Battistoni:2005pd}, placing the recommended 20\% theoretical uncertainty. The final exposures of experiments like DARWIN~\cite{Aalbers:2016jon} and Argo~\cite{Sanfilippo:2019amq}, may reach the high mass section of the neutrino floor set by this background. Since many WIMP-like models with viable cosmologies (see e.g. Refs.~\cite{Athron:2017qdc,Athron:2017ard,Beskidt:2017xsd,Roszkowski:2014iqa,Athron:2017qdc,Hisano:2011cs,Kobakhidze:2018vuy,Arcadi:2017wqi,Baker:2019ndr,Arina:2019tib} for a truncated sample) populate the neutrino fog in this regime, it is a critical part to try to reach.

\textbf{\textit{Statistics}}.---We quantify the impact of the neutrino background in terms of discovery limits. We compute these using the standard profile likelihood ratio test~\cite{Cowan:2010js}, which we will now briefly discuss.

Our parameters of interest are the DM mass and cross section, as well some nuisance parameters in the form of the neutrino flux normalisations $\boldsymbol{\Phi} = \{ \Phi^1, ..., \Phi^{n_\nu} \}$. Since the background-only model should be insensitive to $m_\chi$, we can take it as a fixed parameter and scan over a range of values. We use a binned likelihood written as the product of the Poisson probability in each bin, multiplied by Gaussian likelihood functions for the uncertainties on each neutrino flux normalisation:
\begin{equation}\label{eq:likelihood}
 \mathscr{L}(\sigma,\boldsymbol{\Phi}) = \prod_{i=1}^{N_\textrm{bins}} \mathscr{P} \left[N_\textrm{obs}^i \bigg| N^i_\chi + \sum_{j=1}^{n_\nu} N^{i}_\nu(\Phi^j)\right] \prod_{j=1}^{n_\nu} \mathscr{G}(\Phi^j) \, .
\end{equation}
The Gaussian distributions have standard deviations given by the systematic uncertainties that we just discussed. The Poisson probabilities at the $i$th bin are taken for an observed number of events $N_\textrm{obs}^i$, given an expected number of signal events $N_\chi^i$ and the sum of the expected number of neutrino events for each flux $N_\nu^i(\Phi^j)$. We bin events logarithmically between $10^{-4}$ and 150 keV. The former threshold is clearly not in any way realistic, however the main advantage of our definition of the neutrino floor will be that it is not based on absolute numbers of events. Our choice of threshold is simply to allow us to map the neutrino floor down to $m_\chi = 0.1$~GeV, but crucially it does not impact the height of the limit at other masses.

If we have two models, a background-only model $\mathcal{M}_{\sigma=0}$, and a signal+background model $\mathcal{M}$, we can test for $\sigma > 0$ using the following statistic,
\begin{equation}
	q_0 = \left\{ \begin{array}{rl}
	-2\ln\bigg[\frac{ \mathscr{L} (0,\hat{\hat{\boldsymbol{\Phi}}} | \mathcal{M}_{\sigma=0}) }{\mathscr{L} (\hat{\sigma},\hat{\boldsymbol{\Phi}} | \mathcal{M})  }\bigg]  & \, \,  \hat{\sigma}>0  \, \\
	0  & \, \, \hat{\sigma}\le 0, \, \,  \, 
	\end{array} \right. 
\end{equation}
where~$\mathscr{L}$ is maximised at $\hat{\hat{\boldsymbol{\Phi}}}$ when $\sigma$ is set to 0, and $(\hat{\sigma},\hat{\boldsymbol{\Phi}})$ when $\sigma$ is a free parameter. The model $\mathcal{M}_{\sigma=0}$ is a special case of $\mathcal{M}$, obtained by fixing one parameter to the boundary of its allowed space. Therefore Chernoff's theorem~\cite{Chernoff:1954eli} holds, and $q_0$ should be asymptotically distributed according to $\frac{1}{2}\chi^2_{1}+\frac{1}{2}\delta(0)$ when $\mathcal{M}$ is true~\cite{Algeri:2020pql}.

Evaluating $q_0$ while avoiding the need to collect many Monte Carlo realisations for every point in the parameter space, we exploit the Asimov dataset~\cite{Cowan:2010js}. This is a hypothetical scenario in which the observation exactly matches the expectation for a given model, i.e. $N^i_{\rm obs} = N^i_{\rm exp}$ for all $i$. The test statistic computed assuming this dataset asymptotes towards the median of the chosen model's $q_0$ distribution~\cite{Cowan:2010js}. In high-statistics analyses such as ours, this turns out to be an extremely good approximation, and has been demonstrated multiple times in similar calculations~\cite{OHare:2016pjy,OHare:2020lva,Gaspert:2021gyj}. We fix $N_{\rm obs}$ to be the expected number of neutrino \emph{and} DM events, and require a threshold test statistic of $q_0>9$. Therefore our limits are defined as the expected 3$\sigma$ discovery limits.

\begin{figure*}
\begin{center}
\includegraphics[trim = 0mm 0mm 0mm 0mm, clip, width=0.49\textwidth]{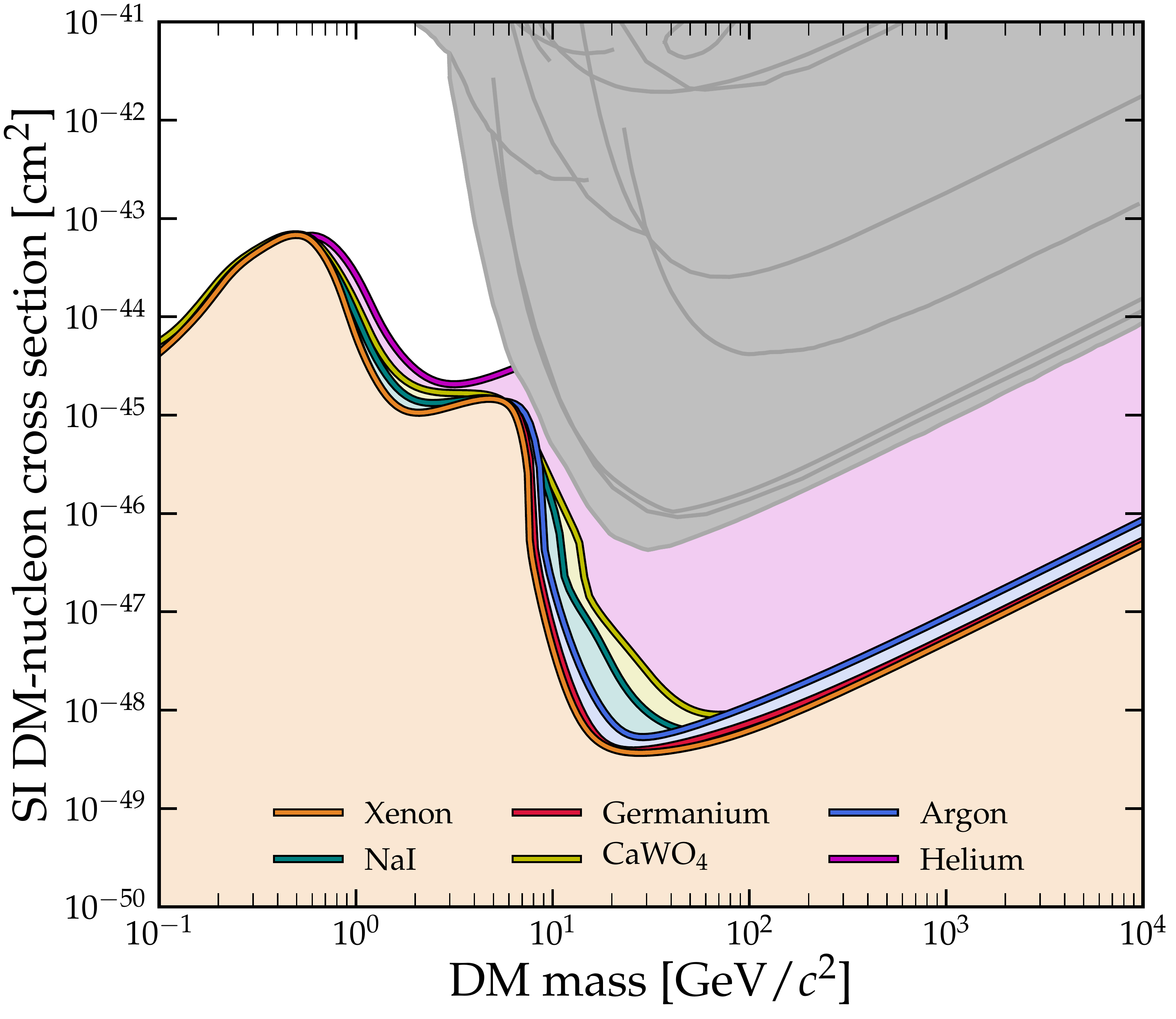}
\includegraphics[trim = 0mm 0mm 0mm 0mm, clip, width=0.49\textwidth]{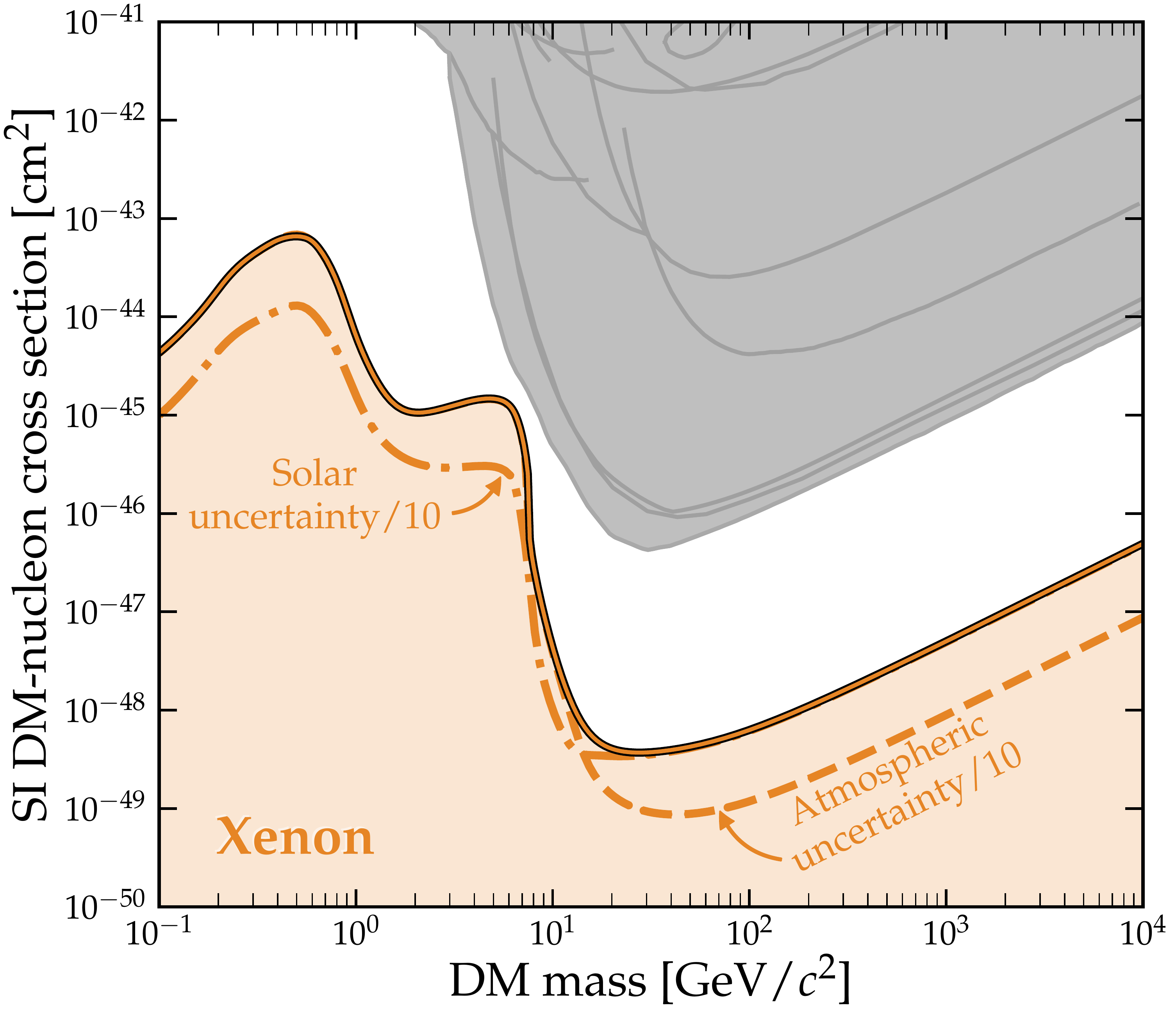}
\caption{{\bf Left:} The neutrino floor for various popular direct detection targets. The general trend is for lighter targets to have floors shifted towards higher masses. The full topology of the neutrino \emph{fog} for these same targets is shown in Fig.~\ref{fig:NuFloor_detailed}. {\bf Right:} Xenon's neutrino floor, showing the effect of improvements to neutrino flux uncertainties. The solid line shows our baseline calculation, whereas the dot-dashed and dashed lines show the result after reducing the systematic uncertainties on the solar and atmospheric fluxes by a factor of 10.} 
\label{fig:NuFloor_Targets}
\end{center}
\end{figure*} 

\textbf{\textit{The neutrino fog}}.---We can now explain how the neutrino background impacts the discovery of DM. The critical factor to understand is the systematic uncertainty on the background. The way to think about this is to imagine a very feeble DM signal that closely matches one of the background components. Such a signal is saturated not just when the number of signal events is simply less than the background, but when that excess of events is smaller than the statistical fluctuations of the background. The regime of parameter space where this occurs is what we define as the \emph{neutrino fog}.

We can quantify the neutrino fog by considering how some discovery limit, $\sigma$, decreases as the exposure/the number of observed background events, $N$, increases. The limit evolves through three distinct scalings. Initially, when the experiment is essentially background free, $N<1$, the limit evolves as $\sigma \propto N^{-1}$. Then, as $N$ increases, the limit transitions into Poissonian background subtraction: $\propto 1/\sqrt{N}$. Eventually, as the number of events increases further, any would-be detectable DM signal disappears beneath the scale of potential background fluctuations, and the limit is stalled at $\sigma \propto \sqrt{(1+N\delta \Phi^2)/N}$~\cite{Billard:2013qya}.

If the DM signal and \cevns background were identical, the saturation regime would persist for arbitrarily large $N$. However there is rarely a value of $m_\chi$ for which the background perfectly matches the signal. In theory one could always collect enough statistics to distinguish the two via their tails or some other broad spectral features. So eventually the limit will emerge from saturation, and the $\sigma\propto N^{-1/2}$ scaling returns, but only by the time $N$ has grown very large. In fact, reaching beyond the saturation regime across the mass range studied here would consume the entire world's supply of atmospheric xenon.

The ``opacity'' of the neutrino fog can therefore be visualised by plotting some gradient of this discovery limit. Let us define the index $n$,
\begin{equation}
    n = -\left( \dbd{\ln \sigma}{\ln N}\right)^{-1} \, ,
\end{equation}
so that $n = 2$ under normal Poissonian subtraction, and $n>2$ when there is saturation (see also Refs.~\cite{Edwards:2017kqw,Edwards:2018lsl,Edwards:2018hcf,Baum:2021jak} for analyses in other contexts that also introduce quantities similar to this). The value of this index for each point in the neutrino fog is shown by the colourscale in Fig.~\ref{fig:NuFloorExplained}. We adopt xenon in this example as it is the most popular target and its corresponding neutrino floor is the most familiar. Seeing as for every value of $m_\chi$ there is a value of $\sigma$ where $n$ crosses 2, we can join these points together in a contour to form the boundary of the neutrino fog, or \emph{neutrino floor}.  

The result of this procedure is the solid line in Fig.~\ref{fig:NewFloor}. There, we compared our result alongside two previous definitions quoted frequently in the literature. The dashed line shows the computation which follows the technique described in Refs.~\cite{Billard:2013qya,Ruppin:2014bra}, with this specific limit taken from Ref.~\cite{Billard:2021uyg}. The technique invoked there involves the interpolation of two limits: a low-threshold/low-exposure one that captures solar neutrinos, and a high-threshold/high-exposure one that excludes solar neutrinos and captures atmospheric and supernovae neutrinos. The two discovery limits are for 400 observed neutrino events so as to put them somewhere around the systematics dominated part of the fog for certain masses.

We also showed a ``one-neutrino'' contour, which is formed from the envelope of a series of background-free 90\% C.L. exclusion limits (2.3 events) with increasing thresholds that have exposures large enough to see one expected neutrino event. These limits have a similar shape but are higher in cross section. One-neutrino contours have an advantage in that they are easy to calculate, but come with the downside that they do not encode any information about the DM/neutrino spectral degeneracy, and do not incorporate systematic uncertainties.

We compare our new neutrino floors for six different targets in Fig.~\ref{fig:NuFloor_Targets}. In the SI space, they are largely similar, with the general trend that the kinematics of scattering off lighter target nuclei means their floors are pushed to higher masses. Helium is the extreme case, because Eq.(\ref{eq:wimprate}) enters the asymptotic $R_\chi\propto\rho_0/m_\chi$ scaling for much lighter masses than other targets, hence why the neutrino floor above 10 GeV is significantly higher, and is set by solar neutrinos rather than atmospheric neutrinos. Molecular targets like NaI and CaWO$_4$, also have some distinctive shape differences due to their multiple nuclei. We show complete maps of the neutrino fog for these targets in Fig.~\ref{fig:NuFloor_detailed}. The broad features observed in these figures largely persist for other interactions, such as spin-dependent scattering. We show examples of these in Fig.~\ref{fig:NuFloor_SD}.

Finally, in the right-hand panel of Fig.~\ref{fig:NuFloor_Targets} we show how our definition of the neutrino floor is adaptable to improvements in the flux estimates. The dashed line imagines that we have obtained a factor of 10 improvement in the atmospheric neutrino uncertainty, i.e.~down to $\sim$2\%, whereas the dot-dashed line imagines that all solar flux estimates are improved by a factor of 10. These scenarios are intended to be illustrative, rather than in anticipation of any specific improvements. Nevertheless, with a flock of large-scale neutrino observatories on the horizon~\cite{Theia:2019non,EricaCadenfortheSNO+:2017rzv,JinpingNeutrinoExperimentgroup:2016nol,Abi:2018dnh,An:2015jdp}, it is not unreasonable to expect some improvement to our knowledge of the neutrino fluxes, especially the poorer measured solar fluxes or the low-energy atmospheric flux~\cite{Cocco:2004ac,Li:2017zix,Capozzi:2018dat,Zhu:2018rwc,Kelly:2019itm,OHare:2020lva}.

\textbf{\textit{Discussion}}.---Given the imminent arrival of the neutrino background in underground WIMP-like DM searches, we have decided to revisit and refine the neutrino floor, or perhaps more appropriately, neutrino fog. Our floor can be interpreted as the \emph{boundary} of the neutrino fog in a statistically meaningful way. It marks the point at which any experiment will start to be limited by the background: a cross section that is influenced only by overlap between the DM and neutrino spectra, and the background's systematic uncertainties.

In contrast to prior calculations, our definition of the neutrino floor is not based on arbitrary absolute numbers of events, but on the \emph{derivative} of the discovery limit. As such we arrive at a limit that does not depend upon the recoil energy threshold, so long as one does not attempt to calculate the limit for masses that only scatter below the chosen threshold. This also means that we do not need to interpolate multiple limits together to map the floor across a wider mass range, and can do so with a single calculation.

The main quantitative differences between our result and those found in the literature are the following. Firstly, we can see for instance in Fig.~\ref{fig:NewFloor}, that our neutrino floor is noticeably higher in the sub-GeV region where $^7$Be neutrinos mimic the signal, as well as the high mass region where the same is true of atmospheric neutrinos. Previous calculations that used fixed exposures, ended up placing the neutrino floor deeper into the systematics dominated regime for those masses compared to others. On the other hand, the shoulder in the neutrino floor around 6 GeV is lower in our case. This is because we have adopted a systematic uncertainty on the \Boron flux of 2\% which is based on a global fit of neutrino data~\cite{Bergstrom:2016cbh}. The previous calculation of the neutrino floor assumed a 15\% systematic uncertainty coming from the solar model estimation.

We have applied our new technique to map the neutrino fog for a range of targets (Fig.~\ref{fig:NuFloor_detailed}) as well as a few different DM interactions (Fig.~\ref{fig:NuFloor_SD}). In the right-hand panel of Fig.~\ref{fig:NuFloor_Targets} we also showed how our neutrino floor can be updated in the future as neutrino observatories make continued refinements to measured fluxes. Our technique is such that these improvements can be incorporated without changing the definition of the neutrino floor. To accompany this paper, we have supplied a public code~\cite{NeutrinoFogCode} with the tools needed to compute the neutrino floor and fog in the manner detailed here.

\textbf{\textit{Acknowledgements}}.---This work was supported by The University of Sydney

\bibliography{neutrinos.bib}
\bibliographystyle{bibi}

\clearpage

\onecolumngrid
\begin{center}
  \textbf{\large Supplementary Figures}\\[.2cm]
  \vspace{0.05in}
  {Ciaran A.~J.~O'Hare}
\end{center}

\onecolumngrid
\setcounter{equation}{0}
\setcounter{figure}{0}
\setcounter{table}{0}
\setcounter{section}{0}
\makeatletter
\renewcommand{\theequation}{S\arabic{equation}}
\renewcommand{\thefigure}{S\arabic{figure}}
\renewcommand{\theHfigure}{S\arabic{figure}}
\renewcommand{\thetable}{S\arabic{table}}

\begin{figure}[htb]
\begin{center}
\includegraphics[trim = 0mm 0mm 0mm 0mm, clip, height=0.3\textwidth]{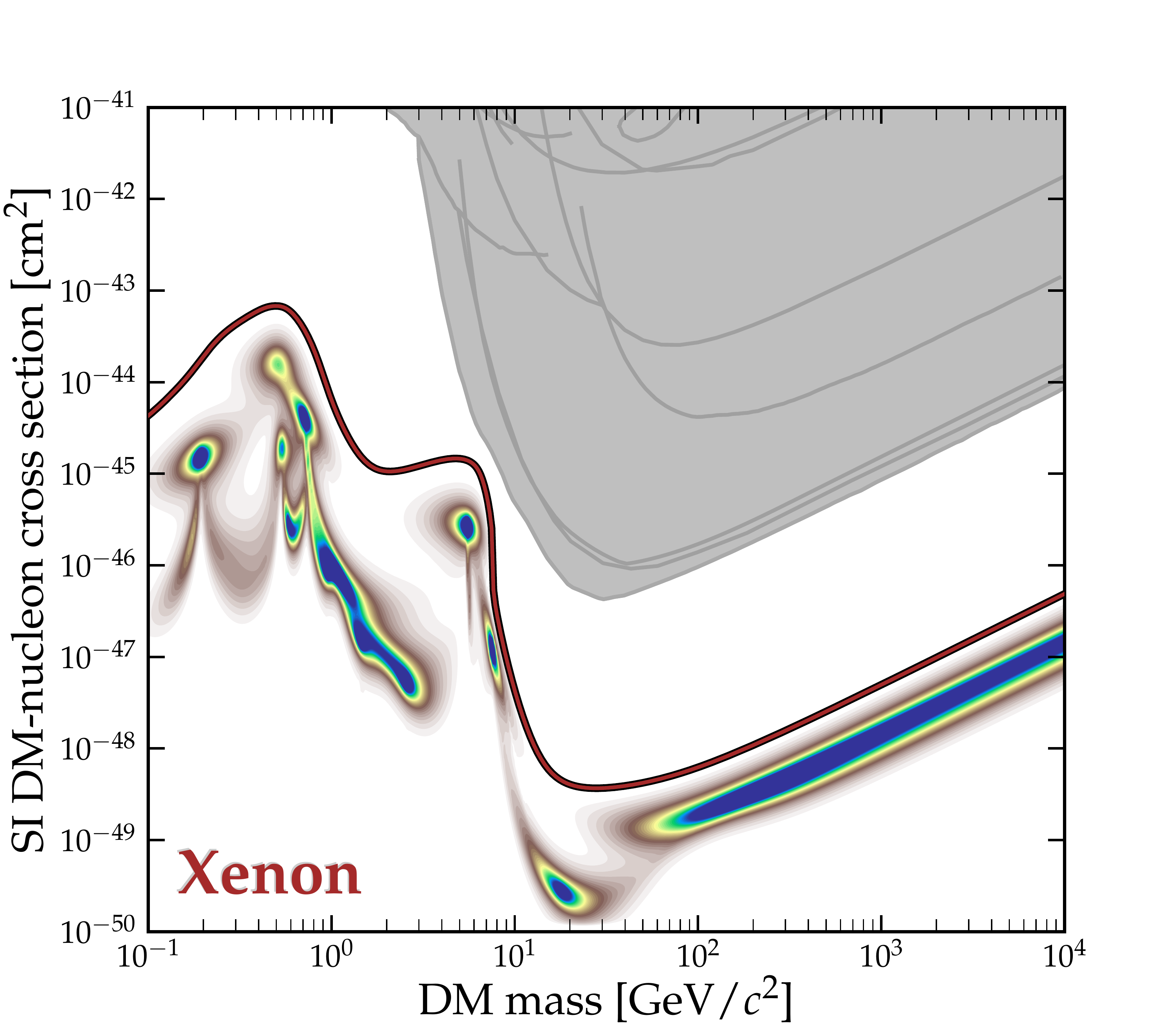}
\includegraphics[trim = 0mm 0mm 0mm 0mm, clip, height=0.3\textwidth]{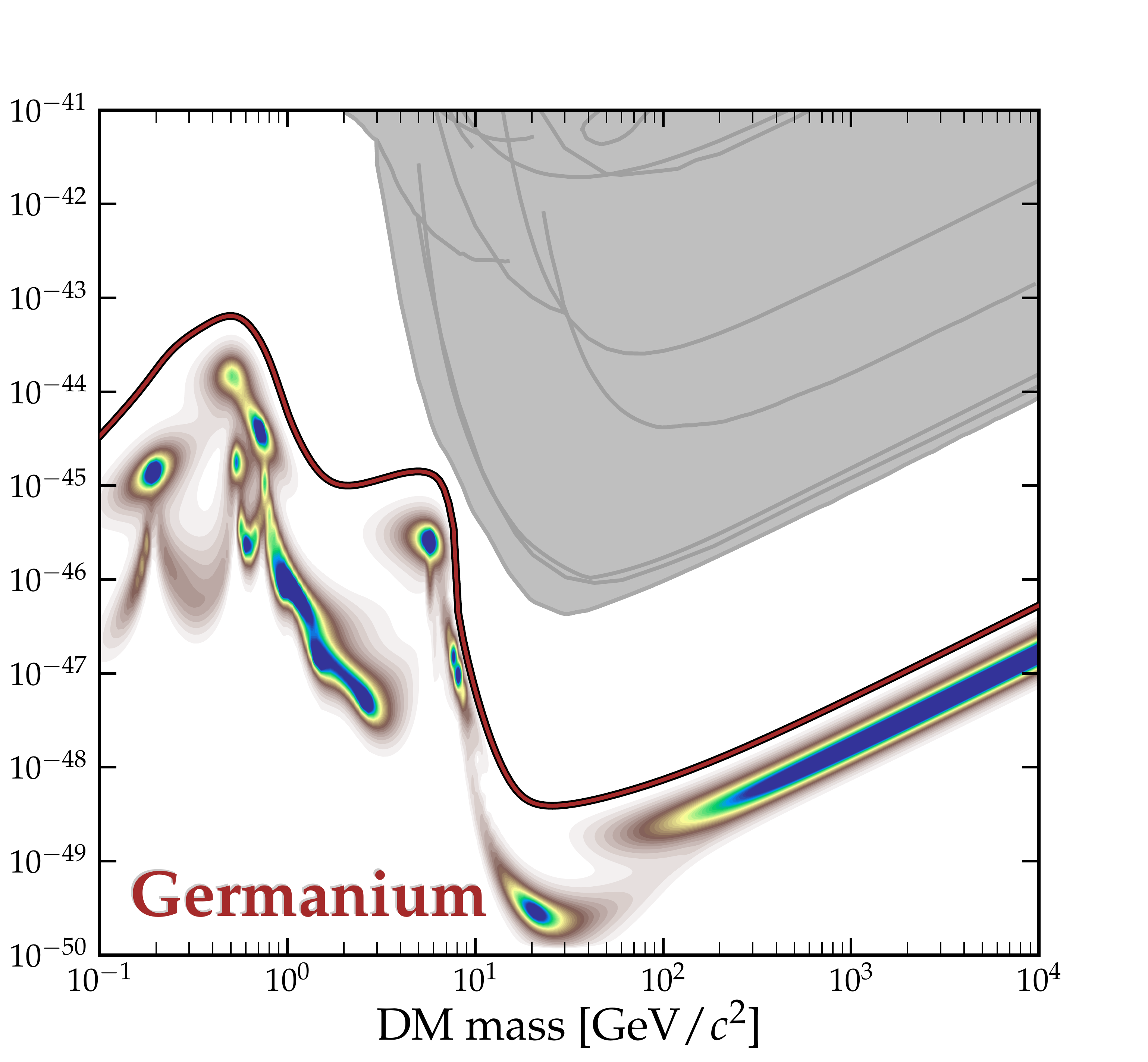}
\includegraphics[trim = 0mm 0mm 0mm 0mm, clip, height=0.3\textwidth]{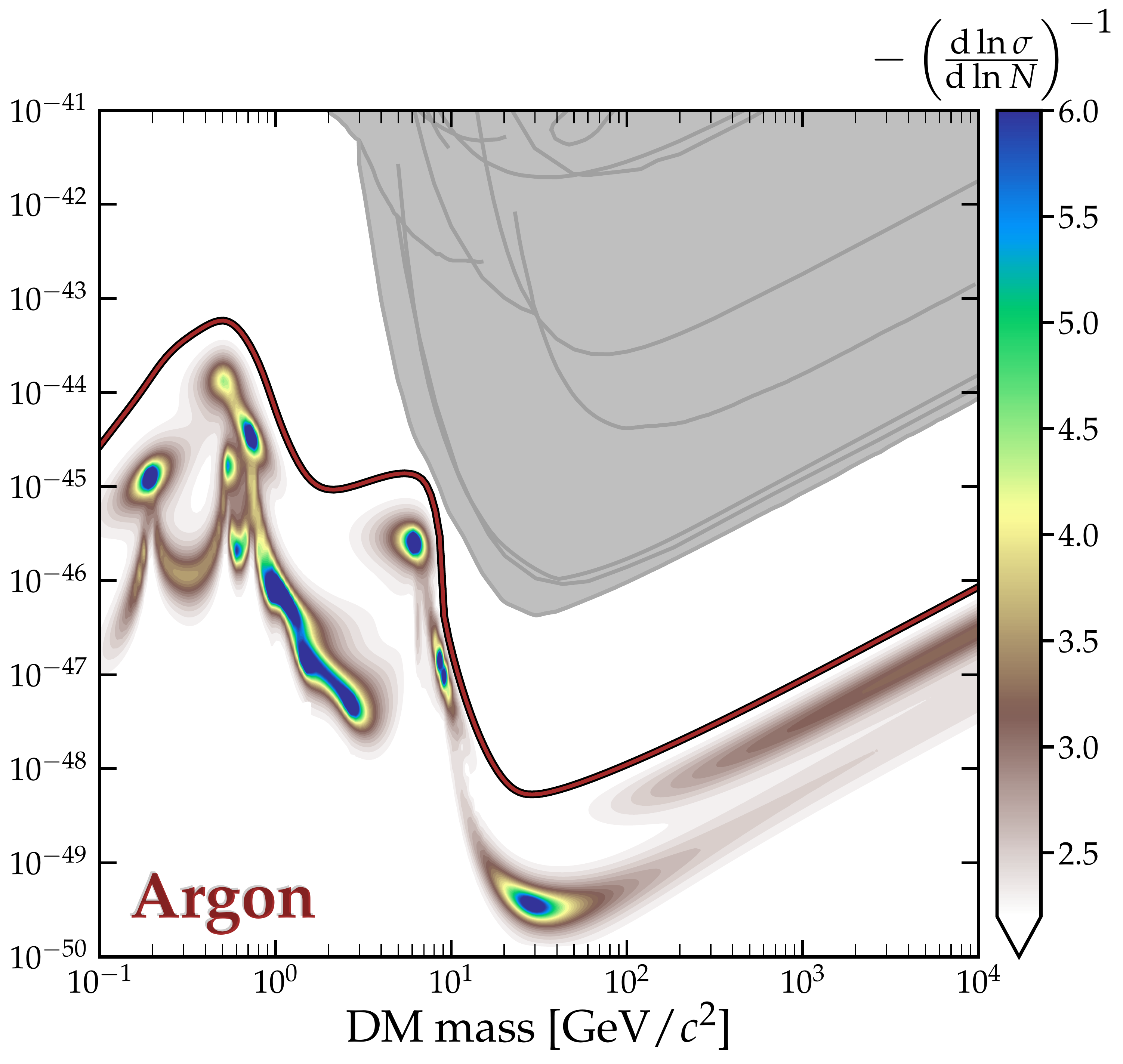}
\includegraphics[trim = 0mm 0mm 0mm 0mm, clip, height=0.3\textwidth]{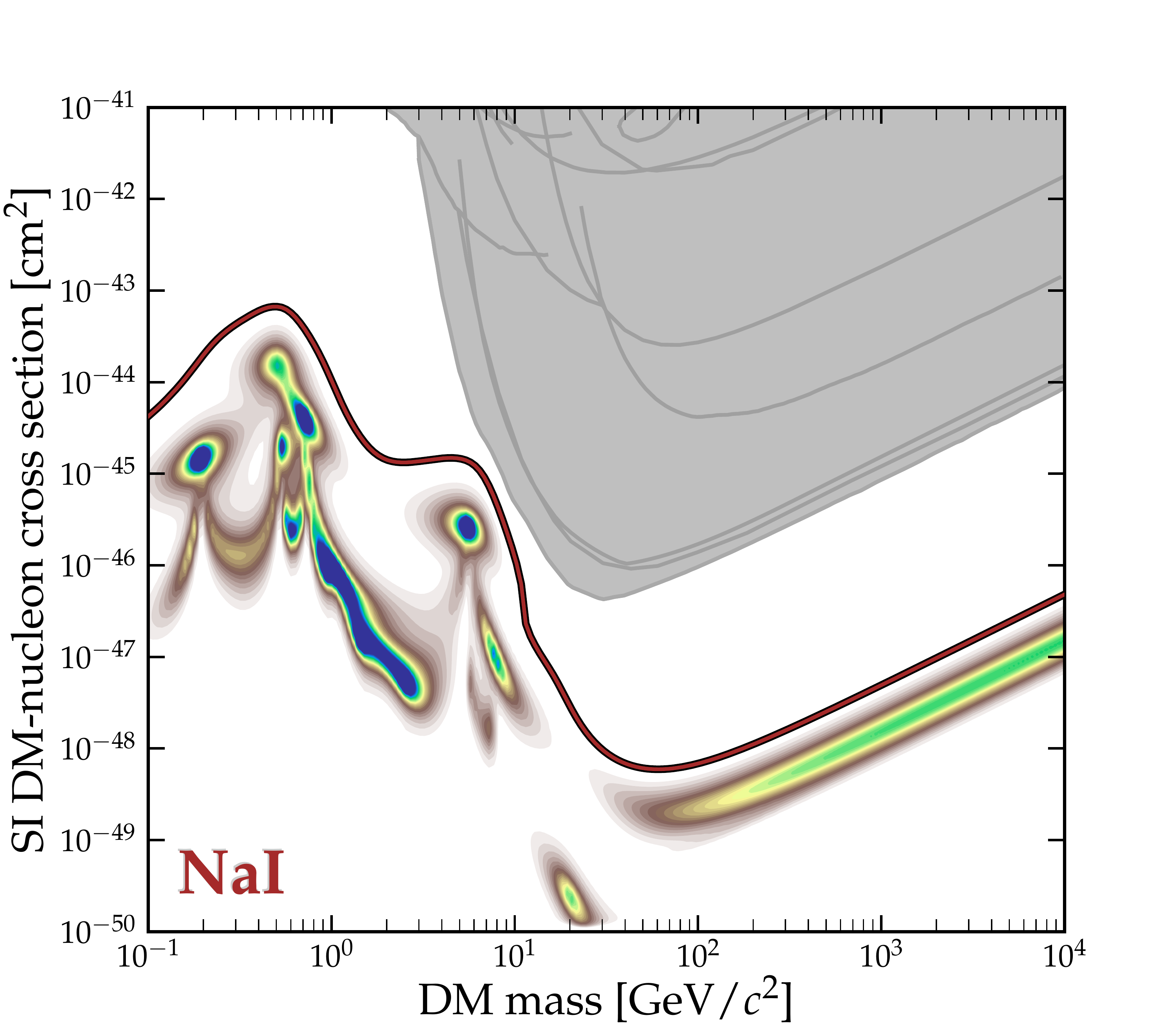}
\includegraphics[trim = 0mm 0mm 0mm 0mm, clip, height=0.3\textwidth]{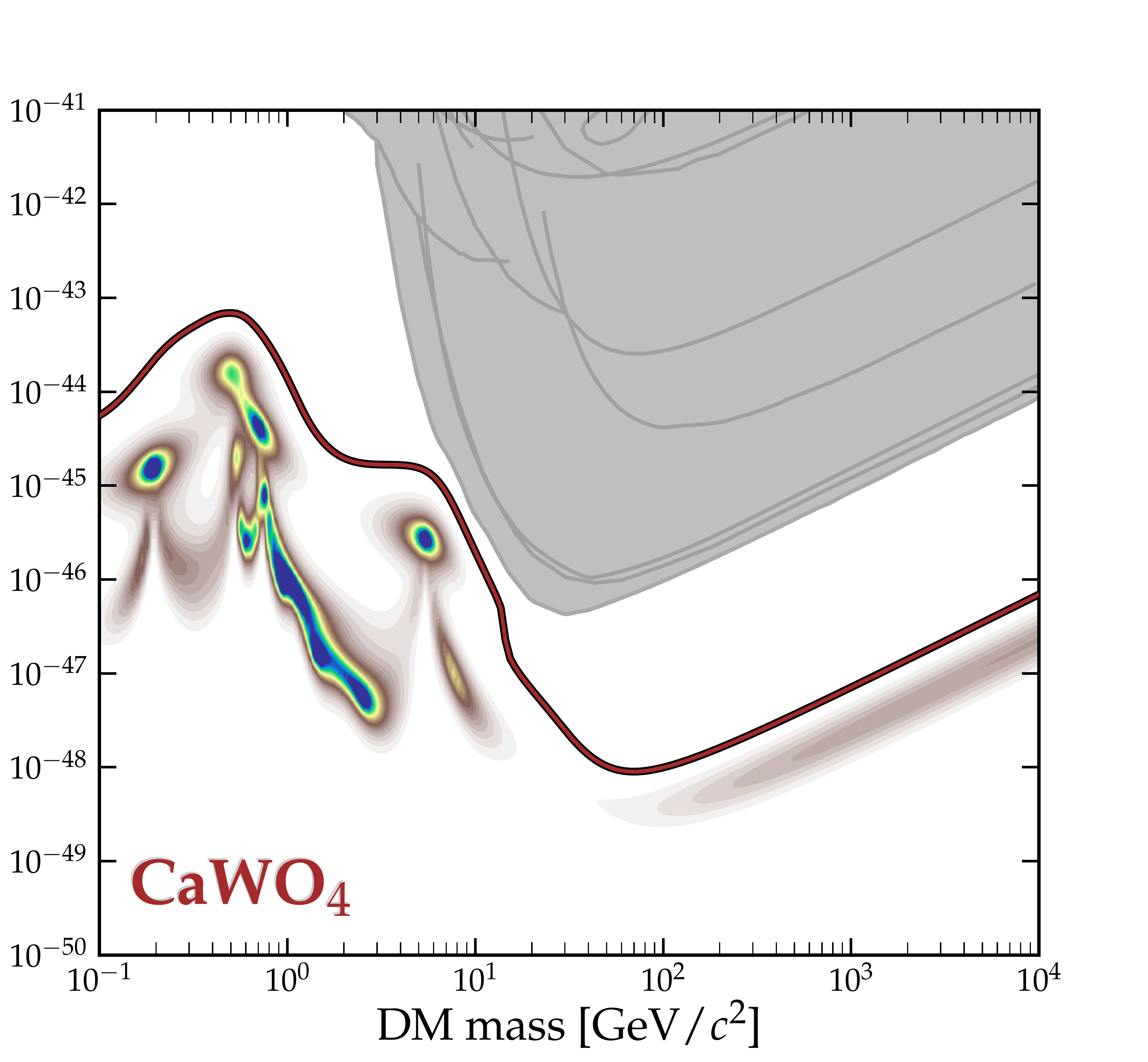}
\includegraphics[trim = 0mm 0mm 0mm 0mm, clip, height=0.3\textwidth]{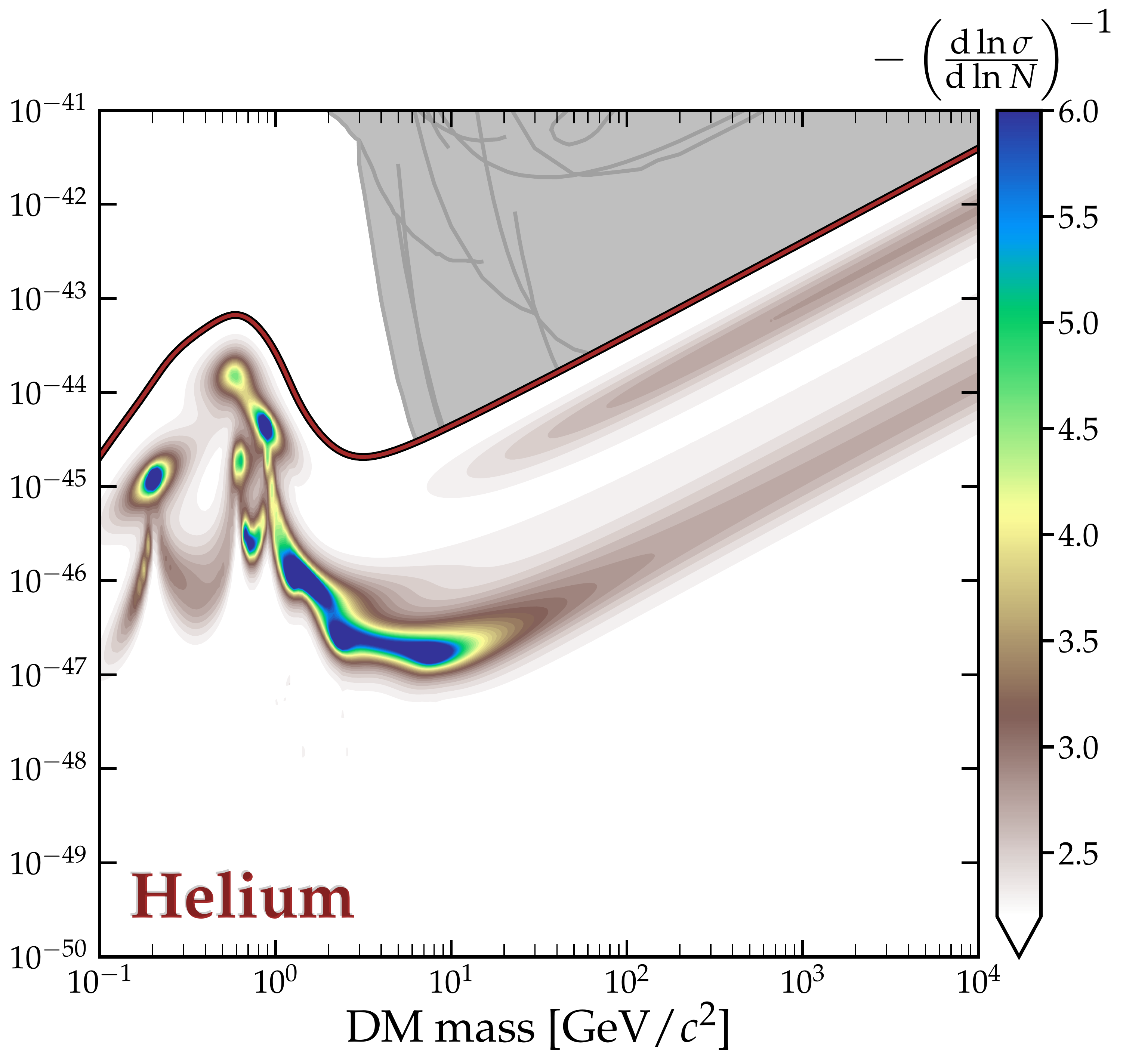}
\caption{Detailed maps of the neutrino fog for six popular direct detection targets: Xe, Ge, Ar, NaI, CaWO$_4$ and He, chosen to provide a wide range of nuclei masses. Notable differences between the impact of the neutrino background on different targets can be noted using the ``opacity'' of the neutrino fog encoded by the value of $n$. In other words, the colourscale displays how similar the neutrino and DM signals are for certain targets, and thus how strongly DM discovery is impacted by the neutrino background. The bright blue colour of the atmospheric region for xenon for instance can be contrasted against the same region in the Ar, CaWO$_4$, and He panels. This shows that the $m_\chi = $100 GeV--10~TeV signal in those targets is notably less well mimicked by atmospheric neutrinos than in Xe or Ge.} 
\label{fig:NuFloor_detailed}
\end{center}
\end{figure} 

\begin{figure}[htb]
\begin{center}
\includegraphics[trim = 0mm 0mm 0mm 0mm, clip, height=0.42\textwidth]{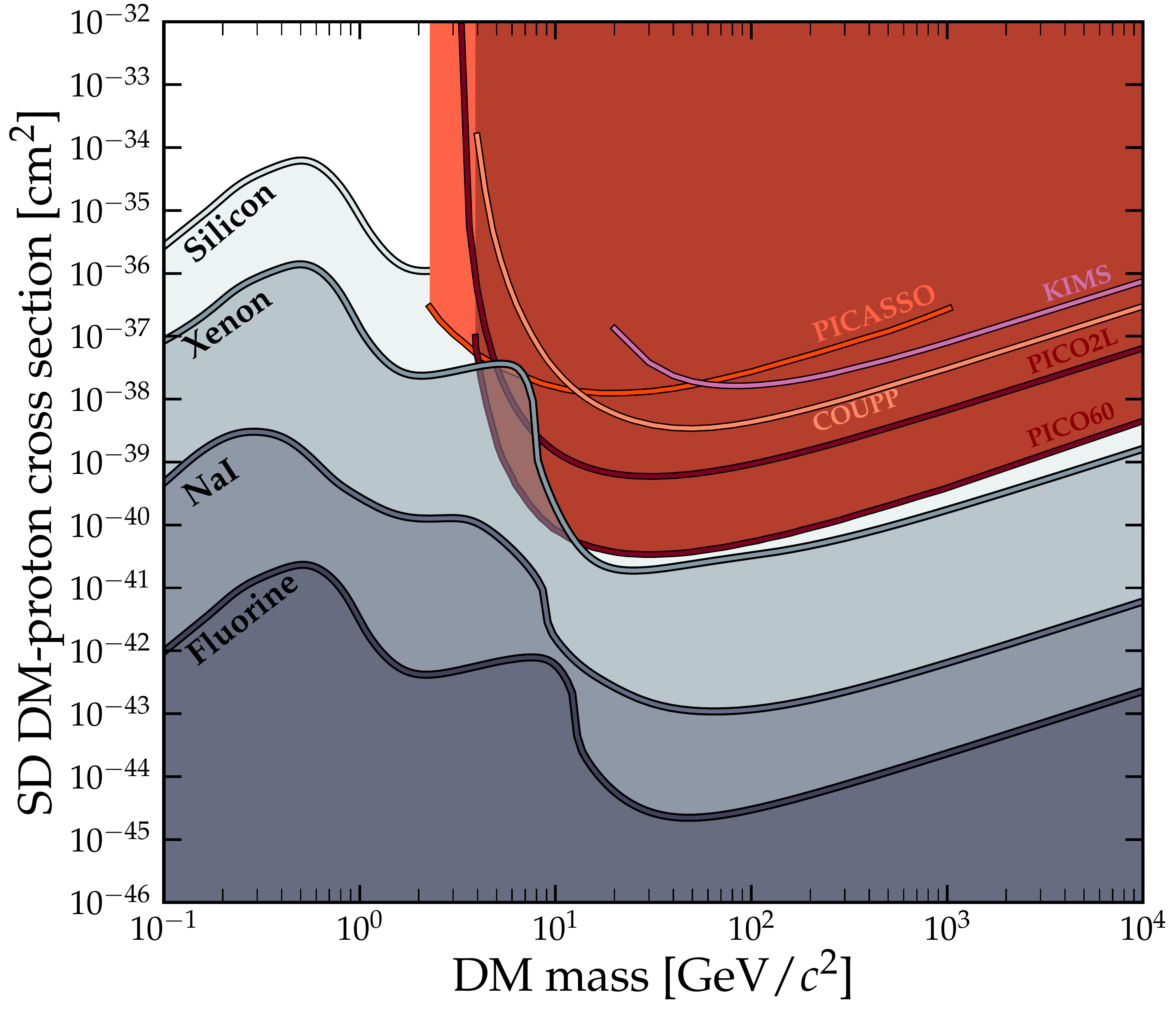}
\includegraphics[trim = 0mm 0mm 0mm 0mm, clip, height=0.42\textwidth]{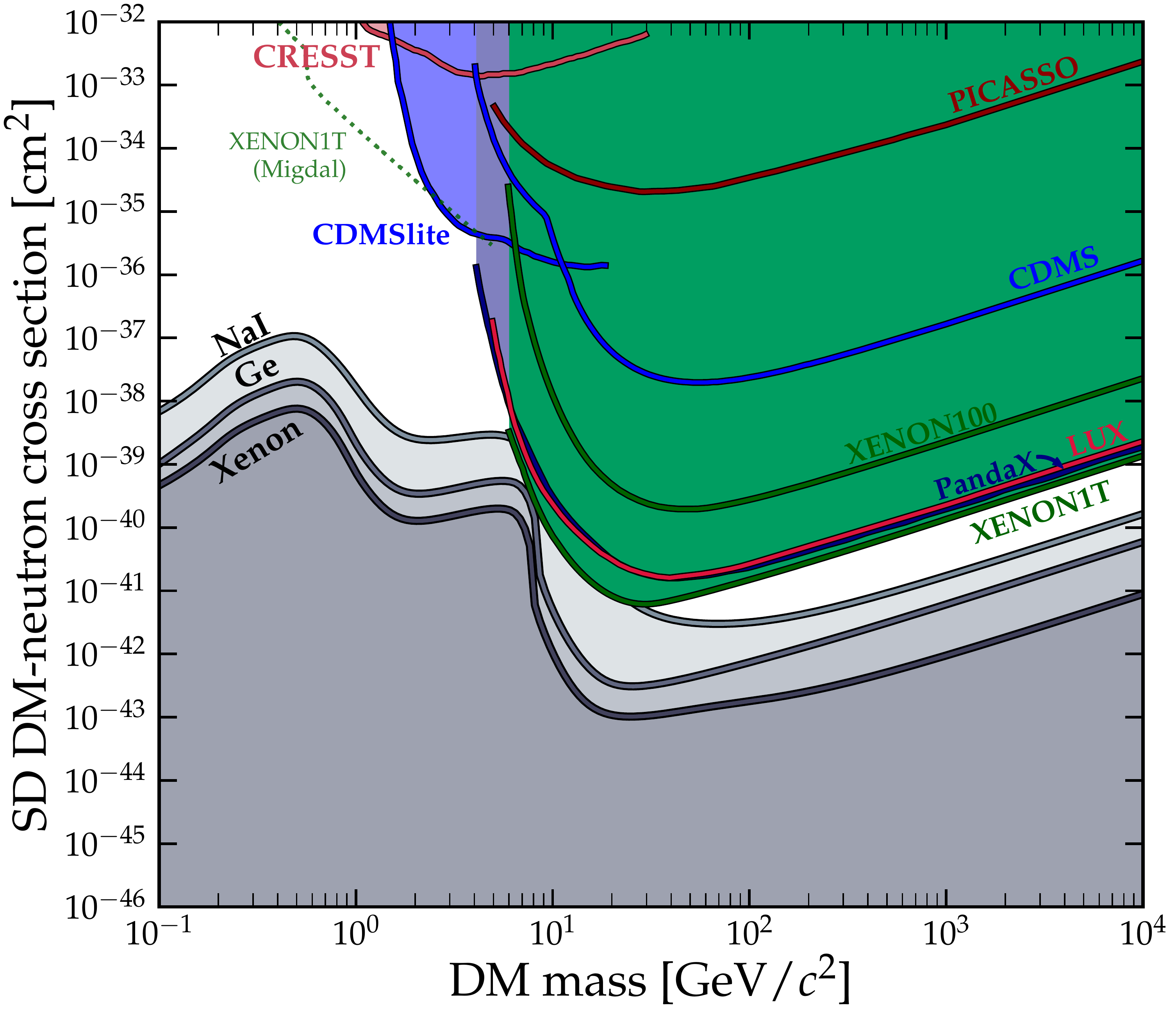}
\caption{Spin-dependent (SD) neutrino floors for various targets with nuclear spin: Si, NaI, Ge, Xe, and F. Exclusion limits are taken from Refs.~\cite{Behnke:2012ys,Kim:2018wcl,Archambault:2012pm,Amole:2016pye, Amole:2017dex} (proton, left) and Refs.~\cite{CRESST:2019jnq,Archambault:2012pm,CDMS-II:2009ktb,LUX:2017ree,PandaX-II:2018woa,XENON100:2016sjq,Aprile:2019jmx,XENON:2019rxp} (neutron, right). The height of the neutrino floor varies much more widely in cross section due to the range of values of the nucleon spin expectation value $\langle S_{p,n}\rangle$ for each target, as well as the fraction of the target comprised of a spin-possessing isotope. For instance, for SD-proton interactions, the fluorine floor is the lowest because the proton spin expectation value of $^{19}$F is high $\sim 0.48$~\cite{Tovey:2000mm}. On the other hand, for SD-neutron interactions, xenon has the highest combination of $\langle S_n\rangle$ on its two spin-possessing isotopes $^{129}$Xe and $^{131}$Xe, which make up around half of natural xenon. We show only a few examples here to reduce clutter, plots for all target nuclei considered can be found at~\cite{NeutrinoFogCode}.} 
\label{fig:NuFloor_SD}
\end{center}
\end{figure} 

\clearpage

\end{document}